\documentclass[twocolumn,aps,prl,superscriptaddress,longbibliography]{revtex4-2}
\usepackage{amsmath}
\usepackage{graphicx}
\usepackage{dcolumn}
\usepackage{color}
\usepackage{bm}
\usepackage{braket}
\usepackage{multirow}
\usepackage{makecell}
\usepackage{booktabs}
\usepackage{physics}
\usepackage{color,soul,xcolor}
\definecolor{bleu}{rgb}{0.1,0.2,0.7} 
\usepackage{natbib}
\usepackage[%
  colorlinks=true,
  urlcolor=bleu,
  linkcolor=bleu,
  citecolor=bleu
]{hyperref}

\begin{document}
\title{Dissipation Driven Coherent Dynamics Observed in Bose-Einstein Condensates}

%%%%%% authors
\author{Ye Tian}
\thanks{These authors contributed equally.}
\affiliation{Department of Physics and State Key Laboratory of Low Dimensional Quantum Physics, Tsinghua University, Beijing, 100084, China}
\author{Yajuan Zhao}
\thanks{These authors contributed equally.}
\affiliation{Department of Physics and State Key Laboratory of Low Dimensional Quantum Physics, Tsinghua University, Beijing, 100084, China}
\author{Yue Wu}
\thanks{These authors contributed equally.}
\affiliation{Institute for Advanced Study, Tsinghua University, Beijing, 100084, China}
\author{Jilai Ye}
\affiliation{Department of Physics and State Key Laboratory of Low Dimensional Quantum Physics, Tsinghua University, Beijing, 100084, China}
\author{Shuyao Mei}
\affiliation{Department of Physics and State Key Laboratory of Low Dimensional Quantum Physics, Tsinghua University, Beijing, 100084, China}
\author{Zhihao Chi}
\affiliation{Department of Physics and State Key Laboratory of Low Dimensional Quantum Physics, Tsinghua University, Beijing, 100084, China}
\author{Tian Tian}
\affiliation{Department of Physics and State Key Laboratory of Low Dimensional Quantum Physics, Tsinghua University, Beijing, 100084, China}
\author{Ce Wang}
\affiliation{School of Physics Science and Engineering, Tongji University, Shanghai, 200092, China}
\author{Zhe-Yu Shi}
\affiliation{State Key Laboratory of Precision Spectroscopy, East China Normal University, Shanghai 200062, China}
\author{Yu Chen}
\email{ychen@gscaep.ac.cn}
\affiliation{Graduate School of China Academy of Engineering Physics, Beijing, 100193, China}
\author{Jiazhong Hu}
\email{hujiazhong01@ultracold.cn}
\affiliation{Department of Physics and State Key Laboratory of Low Dimensional Quantum Physics, Tsinghua University, Beijing, 100084, China}
\affiliation{Beijing Academy of Quantum Information Sciences, Beijing 100193, China}
\affiliation{Frontier Science Center for Quantum Information and Collaborative Innovation Center of Quantum Matter, Beijing, 100084, China}
\author{Hui Zhai}
\email{hzhai@tsinghua.edu.cn}
\affiliation{Institute for Advanced Study, Tsinghua University, Beijing, 100084, China}
\affiliation{Hefei National Laboratory, Hefei 230088, China}
\author{Wenlan Chen}
\email{cwlaser@ultracold.cn}
\affiliation{Department of Physics and State Key Laboratory of Low Dimensional Quantum Physics, Tsinghua University, Beijing, 100084, China}
\affiliation{Frontier Science Center for Quantum Information and Collaborative Innovation Center of Quantum Matter, Beijing, 100084, China}
\begin{abstract}

We report the first experimental observation of dissipation-driven coherent quantum many-body oscillation, and this oscillation is manifested as the coherent exchange of atoms between the thermal and the condensate components in a three-dimensional partially condensed Bose gas. Firstly, we observe that the dissipation leads to two different atom loss rates between the thermal and the condensate components, such that the thermal fraction increases as dissipation time increases. Therefore, this dissipation process serves as a tool to uniformly ramp up the system's temperature without introducing extra density excitation. Subsequently, a coherent pair exchange of atoms between the thermal and the condensate components occurs, resulting in coherent oscillation of atom numbers in both components. This oscillation, permanently embedded in the atom loss process, is revealed clearly when we inset a duration of dissipation-free evolution into the entire dynamics, manifested as an oscillation of total atom number at the end. Finally, we also present a theoretical calculation to support this physical mechanism, which simultaneously includes dissipation, interaction, finite temperature, and harmonic trap effects. Our work introduces a highly controllable dissipation as a new tool to control quantum many-body dynamics. 

\end{abstract}

% make title
\date{\today}
\maketitle

Coherent quantum dynamics, manifested as long-lasting single-frequency-dominated oscillation, has been widely observed in various quantum many-body systems of ultracold atomic gases. The most well-known examples include collective modes of atomic Bose-Einstein condensate \cite{mewes1996collective,stringari1996collective,jin1996collective,jin1997temperature,stamper-kurn98,chevy2002transverse,dalfovo99,onofrio2000surface}, Josephson oscillation between two linked Bose condensates or fermion superfluids \cite{stringari96b,leggett98,cataliotti2001josephson,levy2007ac,mukhopadhyay2024observation,valtolina2015josephson,sukhatme2001observation}, and coherent driven polaron excitation \cite{scazza2017repulsive,oppong2019observation,kohstall2012metastability}. Some recently discovered more intriguing quantum many-body phenomena, such as quantum many-body scars \cite{bernien2017probing,turner2018weak,su2023observation,serbyn2021quantum,wang2023interrelated} and discrete time crystals \cite{zhang2017observation,choi2017observation,yao2017discrete,rovny2018observation,kessler2021observation}, also experimentally manifest as coherent quantum dynamics. These dynamic processes share the common underlying feature that very few well-defined elementary excitations, or many-body eigenstates, dominate the entire quantum dynamics, and the system maintains phase coherence during the dynamics. 

These coherent dynamics are usually excited by a sudden quench or periodical driving a physical parameter in the Hamiltonian because these two protocols can selectively excite a few collective modes that dominate subsequent dynamics. So far, to the best of our knowledge, coherent quantum dynamics driven by dissipation have not been observed before. Although no fundamental obstacle prevents coherent dynamics from occurring in the dissipation process, it is challenging to observe such an effect because dissipation usually leads to decay or diffusion that conceals coherent dynamics.

In this letter, we report the experimental observation of dissipation-driven coherent dynamics in a finite-temperature Bose-Einstein condensate. This coherent dynamics is driven by applying dissipation inducing the loss of atoms, and this observation takes advantage of the fact that dissipation is highly controllable in ultracold atom systems. First, we find that the same dissipation induces different loss rates of atoms in the condensate and thermal components. This makes dissipation a controllable tool to uniformly and smoothly ramp the system's temperature. Secondly, we can control the strength of dissipation in a time-dependent manner so that we can turn on and off the dissipation at different times. This allows us to temporally separate the coherent dynamics from the decay of atom number because otherwise, the coherent dynamics are always embedded in the loss dynamics and are hard to observe. Thirdly, because the temperature increase is smooth enough in a controllable way and because the harmonic trap imposes an infrared cutoff of the low-energy thermal modes, only a few low-lying thermal modes with a discrete energy spectrum dominate the dynamics. With these three key ingredients, we observe coherent exchange dynamics between the thermal and condensate components. We also provide a theoretical calculation that includes dissipation, interaction, and the harmonic trap, showing that the coherent dynamics are a cooperative effect of these three.

\begin{figure}[t!]
    \centering
    \includegraphics[width=0.48\textwidth]{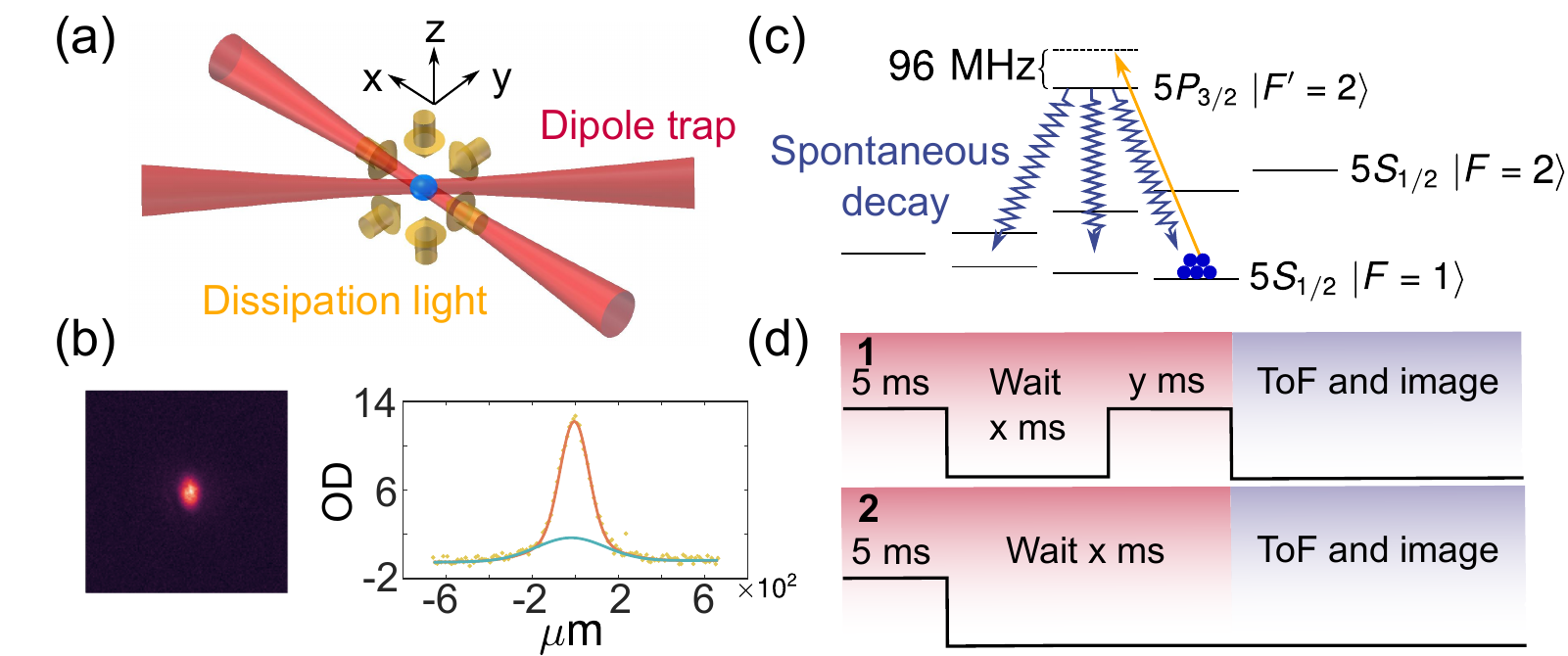}
    \caption{Illustration of the experimental setup and sequence. (a) $^{87}$Rb atoms are loaded into a crossed dipole trap and evaporated to a partially condensed phase. Then, a $96$~MHz blue-detuned dissipation light beams from six directions are applied to the atoms. An absorption imaging is applied along the $x$ axis. (b) Typical time-of-flight (ToF) image of the atoms (left) and their integrated optical density along $z$ axis (right), with which a bimodal fitting yields both condensate and thermal components.  (c) Physical mechanism of the dissipation process. Atoms are initially prepared in $\ket{F=1,m_F=1}$ state, and they will leave the trap if they are pumped out of $\ket{F=1,m_F=1}$  and $\ket{F=2,m_F=-1}$ states by dissipation light. (d) Experimental sequence. For the first protocol denoted as ``$5+x+y$", the dissipation light is first turned on for $5$~ms and then turned off, letting atoms evolve for $x$~ms without dissipation, then turned on again for $y$~ms before ToF measurement. For the second protocol, denoted as ``$5+x$", the dissipation light is turned on for $5$~ms and then turned off, and the ToF measurement is performed right after the dissipation-free evolution of $x$~ms. }
    \label{setup}
\end{figure}

\textit{Experimental System.} In our experiment, we first used a crossed dipole trap formed by $1064$~nm laser to capture $^{87}$Rb atoms polarized to $\ket{F=1,m_F=1}$. Then, these atoms are evaporated to a partially condensed state, where the condensate and thermal fraction are extracted by fitting a bimodal distribution, as shown in Fig. \ref{setup}(b). We vary the condensate fraction by controlling the final trap depth of the evaporative cooling. After evaporation, we adiabatically ramp up the trap depth to a fixed final value, and this process is adiabatic enough such that the condensate fraction does not change.

After preparing a mixture of condensate and thermal atoms, we apply a blue-detuned light that couples atoms to $5P_{3/2}$ states to introduce dissipation. This dissipation light is applied from six counter-propagating directions, as shown in Fig.~\ref{setup}(a). In our experiment, the confinement from the optical dipole trap is relatively shallow and cannot hold atoms against gravity. We have applied a magnetic field gradient to compensate gravity for atoms in the $\ket{F=1,m_F=1}$ and $\ket{F=2,m_F=-1}$ states. Atoms excited to the $5P_{3/2}$ state can decay into eight hyperfine levels $\ket{F=1,m_F=0,\pm 1}$, $\ket{F=2,m_F=0,\pm 1,\pm 2}$ in the ground state manifold, as shown in Fig.~\ref{setup}(c). Among these eight states, atoms returned to $\ket{F=1,m_F=0,-1}$ and $\ket{F=2, m_F=0,1,\pm 2}$ states cannot be held by the optical dipole trap because the gravity cannot be compensated by magnetic field gradient for these six spin states, and atoms in these spin states will be lost from the system while atoms in $\ket{F=1,m_F=1}$ and $\ket{F=2,m_F=-1}$ remain \cite{zhao2023observation}. This is the main dissipation mechanism in our system. For the absorption image, we first pump all atoms to $\ket{F=2,m_F=2}$, so that all the remained atoms can be detected.

\begin{figure}[t!]
    \centering
    \includegraphics[width=0.48\textwidth]{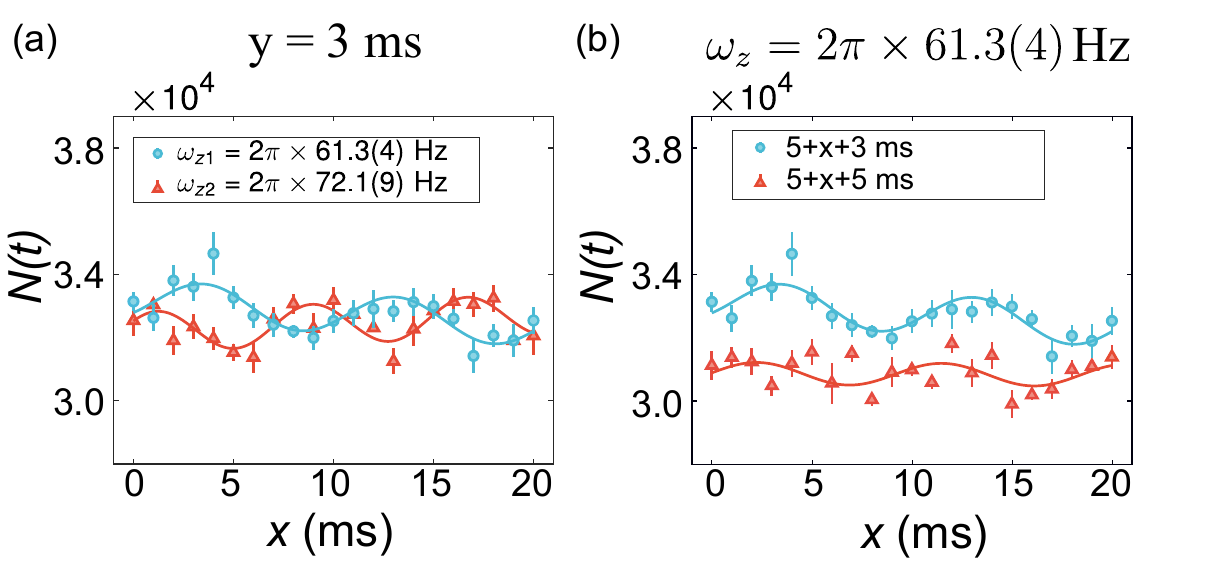}
    \caption{Experimental observation of the dissipation driven coherent dynamics. We start with a partial condensate with the condensate-to-thermal ratio $\sim 2:1$. After the experimental sequence ``$5+x+y$" shown in Fig. \ref{setup}(d), the total number of the remaining atoms oscillates as a function of the waiting duration $x$. (a) shows two data sets with the same experimental sequences but different harmonic trapping frequencies along $\hat{z}$, $\omega_z=2\pi \times 61.3\pm 0.4$~Hz for blue circles and $\omega_z=2\pi\times 72.1\pm 0.9$~Hz for red triangles. Here $y=3$~ms. The fitted frequencies are $2\pi \times 105\pm 6$~Hz for blue circles and $2\pi \times 129\pm 6$~Hz for red triangles. (b) shows two data sets with the same trapping frequency $\omega_z=2\pi \times 61.3\pm 0.4$~Hz but different duration $y$ of the second dissipation process. The fitted frequencies are $2\pi \times 105\pm 6$~Hz and $2\pi \times 108\pm 16$~Hz  for blue circles (5+x+3) and red triangles (5+x+5), respectively. All data points are averaged over nine repeated experiments.}
    \label{oscillation}
\end{figure}

\textit{Main Experimental Finding.} Our main findings are shown in Fig. \ref{oscillation}. To reveal the coherent dynamics during dissipation, we design a specific experimental protocol in which a dissipation-free evolution is inserted between two periods of dissipation, as shown in the upper sequence in Fig. \ref{setup}(d). After preparing a partially condensed Bose gas, we turn on the dissipation light for $5$~ms to introduce atom loss. Then, we turn off the dissipation and let the entire system evolve for a duration of $x$~ms. We have verified that the total number of atoms remains constant during this waiting period. Then, we turn on the same dissipation light again for $y$~ms to further introduce atom loss, after which we measure the remaining atoms using the ToF imaging. Surprisingly, we find that the number of remaining atoms after the entire sequence oscillates as a function of waiting duration $x$, as shown in Fig. \ref{oscillation}. 

Fig. \ref{oscillation}(a) shows two typical oscillations for atoms trapped in two different trap frequencies. One can see that the oscillation frequencies of the total atom number vary as the trap frequency changes, but the oscillation amplitudes are nearly the same. Fig. \ref{oscillation}(b) shows two typical oscillations with the same trap frequency but two different durations $y$ of the second dissipation period. The oscillation frequencies are nearly the same for these two cases, but the oscillation amplitudes gradually diminish as $y$ increases. This is natural because a long enough dissipation will eventually smear out coherent signatures. We have varied trap frequencies, initial condensate fractions, and different dissipation durations and have found such oscillations are universal behaviors. We have also confined the system into a one-dimensional tube by applying a two-dimensional optical lattice, and similar phenomena have also been observed. 

\begin{figure}[t]
    \centering
    \includegraphics[width=0.48\textwidth]{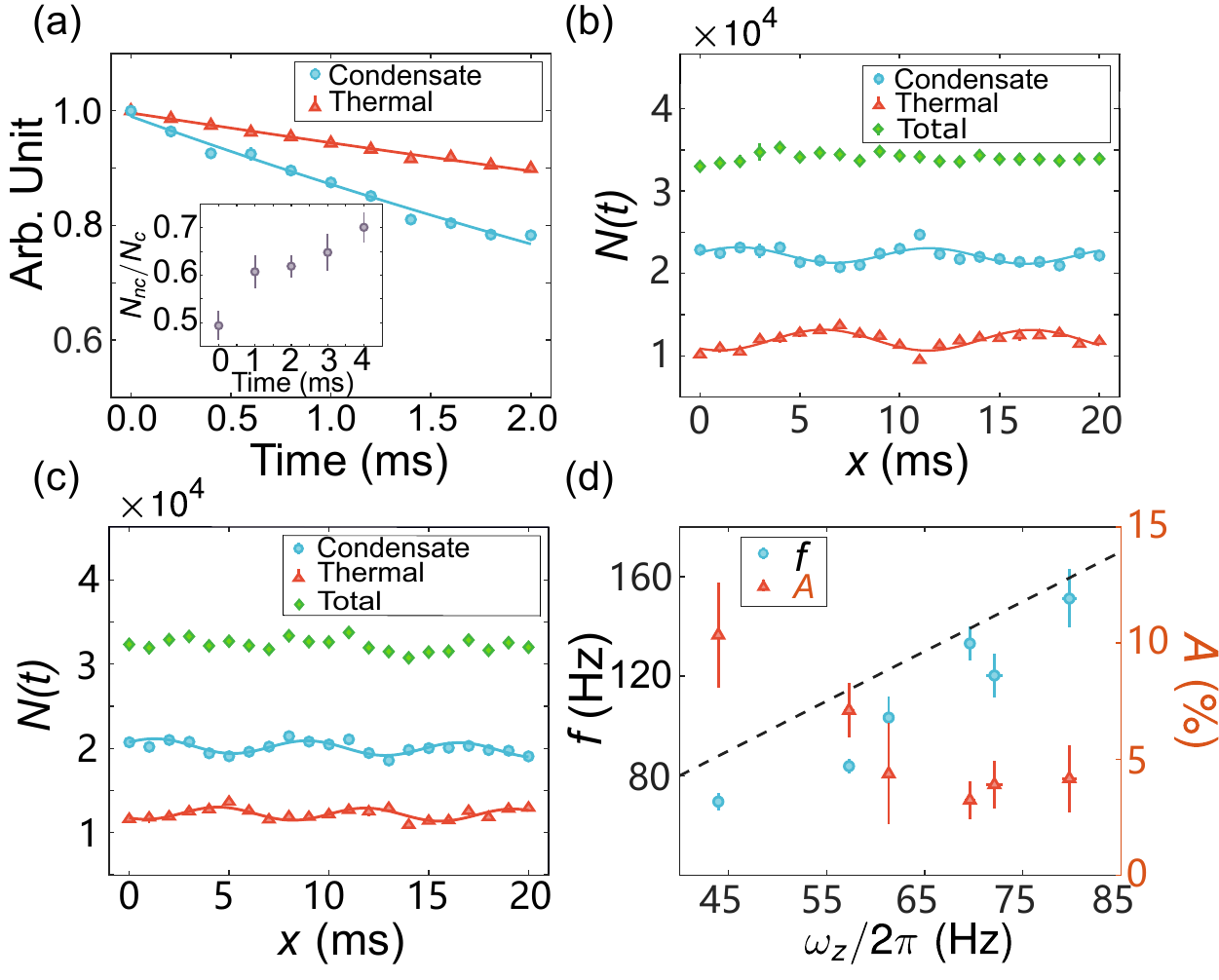}
    \caption{(a) Decay of atom number from condensate (blue circles) and thermal component (red triangles) for initially prepared pure condensate or pure thermal gas. The inset shows the increase of the thermal-to-condensate ratio as a function of dissipation time for an initially partially condensed sample. (b-c) Before turning on the second dissipation period, oscillations between the number of condensate atoms and thermal atoms are observed as a function of waiting duration $x$. The blue circles and the red triangles are condensate and thermal atoms, respectively. The green diamonds denote the total number of atoms. $\omega_z=2\pi \times 61.3\pm0.4$~Hz for (b) and $2\pi \times72.1\pm0.9$~Hz for (c). The solid lines are fitting to single-frequency oscillation. (d) The oscillation frequency $f$ (blue circles) and relative amplitude $A$ (red triangles) for different trap frequencies. The dashed line denotes $2\omega_z/2\pi$. All data points are averaged over nine repeated experiments.  }
    \label{mechanism}
\end{figure}

\textit{Physical Mechanism.} Below, we discuss how dissipation drives such coherent dynamics. First, we find that our dissipation mechanism leads to two different loss rates for condensate and thermal components. Experimentally, we prepare a pure condensate or a pure thermal gas and measure their decay rate under the same dissipation strength. The results are shown in Fig. \ref{mechanism}(a). By fitting the decay curves, we obtain a decay time $\tau=7.9 \pm 0.4$~ms for condensate and $\tau=18.6\pm 0.7$~ms for thermal atoms. 
Recall that atoms excited by the dissipation lights can decay into eight hyperfine levels of the ground state, and only atoms returned to $\ket{F=1, m_F=1}$ and $\ket{F=2, m_F=-1}$ states remain trapped. Under this process, a thermal atom remains thermal. However, an atom initially in the zero-momentum condensate can become thermal due to the photon recoil effect in this process. A detailed Clebsch-Gordan coefficients calculation shows that the probability of returning to trapped states versus untrapped states is approximately $0.41:0.59$ \cite{supplementary}, which is consistent with two different decay time scales. As the condensate decays faster, we observe an increase in the thermal-to-condensate ratio as the dissipation time increases, as shown in the inset of Fig. \ref{mechanism}(a).

\begin{figure}[t]
    \centering
    \includegraphics[width=0.48\textwidth]{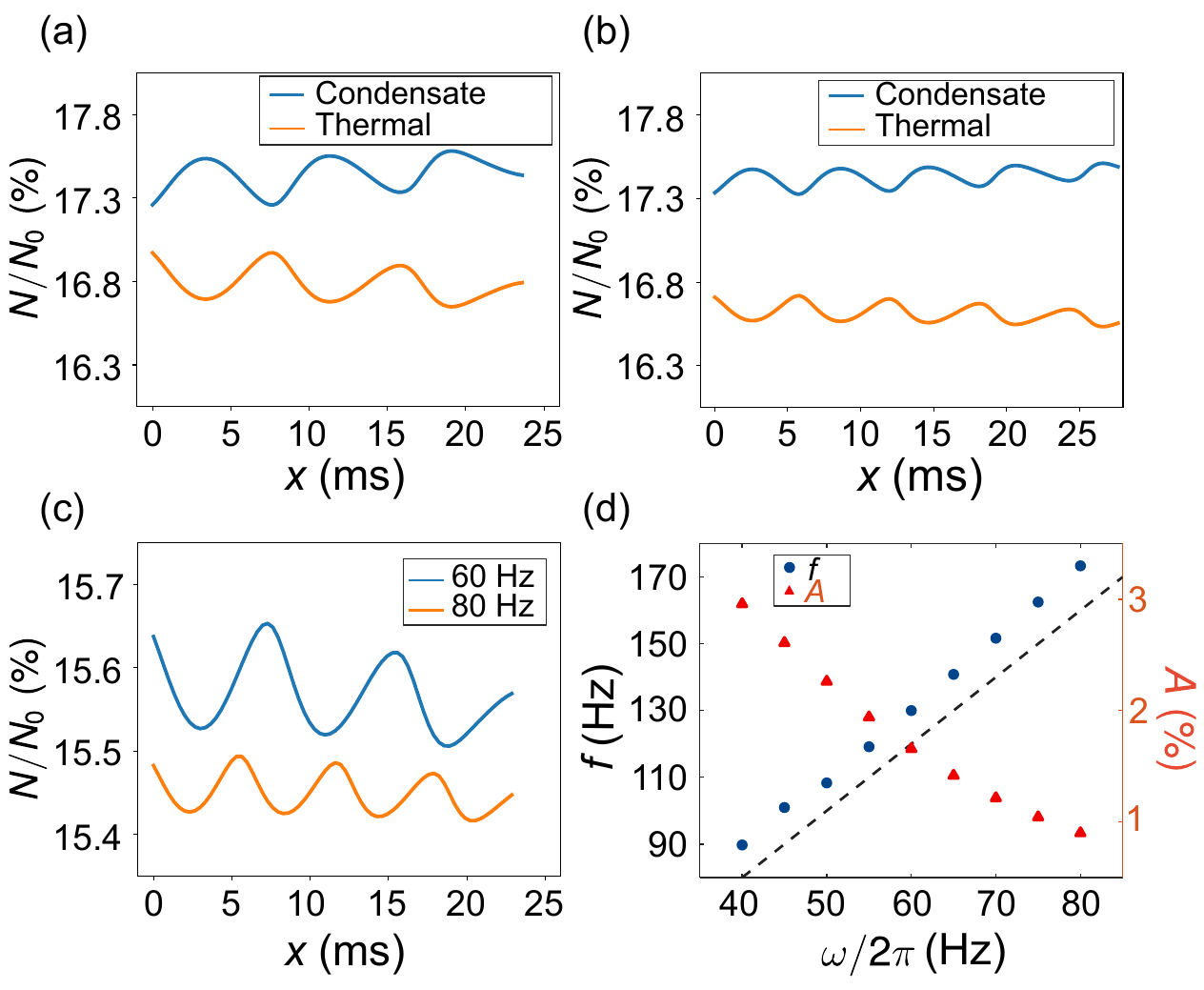}
    \caption{(a-b) Oscillation of the condensate (blue line) and the thermal atoms (orange line) as a function of free evolution time $x$ for $``5+x"$ protocol. Here we choose the trapping frequency $\omega_z=2\pi\times 60$~Hz for (a) and $\omega_z=2\pi\times 80$~Hz for (b). The initial total atom number is $N_0=1033$ for $\omega_z=2\pi\times 60$~Hz and  $N_0=765$ for $\omega_z=2\pi\times 80$~Hz and the condensate fraction is approximately $75\%$ for all cases. The first dissipation strength is taken as $2\omega_z$ for condensate and $\omega_z$ for thermal components, respectively. The interaction strength is chosen as $0.025~\hbar\omega_za_0$ where $a_0=\sqrt{\hbar/m\omega_z}$ is the harmonic length. The first dissipation period is $0.37/\omega_z$. 
(c) The total number of atoms as a function of $x$ after the $``5+x+y"$ protocol, with $y$ fixed at $0.37/\omega_z$. (d) The oscillation frequency $f$ (blue circles) and relative amplitude $A$ (red triangles) for different trap frequencies. The dashed line denotes $f=2\omega_z/2\pi$. }
    \label{theory}
\end{figure}

Since the dissipation changes the condensate fraction and alters the equilibrium between thermal and condensate components, it can drive an exchange dynamics between thermal and condensate components. Fig. \ref{mechanism}(b) and (c) shows that during the exchange dynamics, the thermal and condensate components display an out-of-phase oscillation while the total number remains constant. As explained in the theory part below, these dynamics are driven by pair production of Bogoliubov quasi-particles. Moreover, the presence of a harmonic trap is crucial because it introduces an infrared cut-off of the oscillation frequency; otherwise, all different frequencies mix up, smearing out the coherent oscillation. Fig. \ref{mechanism}(b-c) also shows a general trend that the oscillation frequency increases, and the oscillation amplitude decreases as the trap frequency increases. Roughly speaking, the oscillation frequencies follow the trend of $2\omega_z/2\pi$ as shown in Fig. \ref{mechanism}(d). This is consistent with the property of pair creation and annihilation of Bogoliubov quasi-particles, which is the main process driving the coherent dynamics. 

Hence, the ratio of the thermal and condensate components oscillates as the duration of the waiting time increases. When the second dissipation light is turned on, if the dynamics oscillate to a larger condensed component, the averaged dissipation rate is larger, yielding more atom loss during the second dissipation period, and vice versa. In this way, the coherent dynamics during the free evolution period manifest as the oscillation of the total atom number after the second dissipation period. This explains our main findings.

\textit{Theoretical Calculations.} To further support the physical mechanism discussed above, we present a theoretical calculation including dissipation, interaction, and the harmonic trap on equal footing. First, we implement the standard Bogoliubiv theory to introduce an initial equilibrium state at a finite temperature \cite{supplementary, pethick2008bose, dalfovo99}. We expand the bosonic operator $\hat{\Psi}(x,t)=\Phi(x,t)+\hat{\psi}(x,t)$, where $\Phi(x,t)$ is the condensate wave function and $\hat{\psi}(x,t)$ describes the thermal component. To properly describe the dissipation effect \cite{barontini2013controlling,sels2020thermal,wang2022complex}, we describe the thermal part by the density matrix, including the normal component $n(x^\prime,x,t)=\langle\hat{\psi}^\dag(x^\prime,t)\hat{\psi}(x,t)\rangle$ and the anomalous component $n_\text{s}(x^\prime,x,t)=\langle\hat{\psi}(x^\prime,t)\hat{\psi}(x,t)\rangle$. The anomalous component of the density matrix is essential for capturing the pair creation and annihilation processes. They satisfy a set of coupled equations as follows 
\begin{widetext}
\begin{align}
&i\hbar\frac{\partial \Phi(x,t)}{\partial t}=\left(\hat{H}_0+g|\Phi(x,t)|^2+2g n(x,x,t)\right)\Phi(x,t)+g n_\text{s}(x,x,t))\Phi^*(x,t)-i\gamma_\text{c}\Phi(x,t); \label{GP} \\
&i\frac{\partial n(x^\prime, x,t)}{\partial t}=\left(\hat{H}_0(x)+\mathcal{W}(x)-\hat{H}_0(x^\prime)-\mathcal{W}(x^\prime)\right)n(x^\prime, x,t)+\mathcal{K}(x)n_\text{s}^*(x,x^\prime,t)-\mathcal{K}^*(x^\prime)n_\text{s}(x^\prime,x,t)-i\gamma_\text{t}n(x^\prime, x,t);\label{n}\\
&i\frac{\partial n_\text{s}(x^\prime, x,t)}{\partial t}=\left(\hat{H}_0(x)+\mathcal{W}(x)+\hat{H}_0(x^\prime)+\mathcal{W}(x^\prime)\right)n_\text{s}(x^\prime, x,t)+\mathcal{K}(x)n(x^\prime,x,t)+\mathcal{K}(x^\prime)n(x,x^\prime,t)-i\gamma_\text{t}n_\text{s}(x^\prime, x,t), \label{ns}
\end{align}
\end{widetext}
where $\hat{H}_0=-\frac{\hbar^2}{2m}\frac{\partial^2}{\partial x^2}+\frac{1}{2}m\omega^2 x^2$ is the free particle Hamiltonian. $\mathcal{W}(x)=2g|\Phi(x,t)|^2+2g n(x,x,t)$ and $\mathcal{K}(x)=g\Phi(x,t)^2+gn_\text{s}(x,x,t)$ introduce non-linear coupling between them. $\gamma_\text{c}$ and $\gamma_\text{t}$ represent the dissipation rates for the condensate and the thermal components, respectively. 

The results from the numerical simulation of Eq. \ref{GP}-\ref{ns} are shown in Fig. \ref{theory} \cite{opensource}. Following the experimental protocol, we turn the dissipation on and off at different quantum dynamics stages. We only focus on the one-dimensional situations to reduce the computation load for these coupled equations. The results qualitatively agree with experimental observations but with a much smaller amplitude because of the relatively smaller atom number and weaker interaction effect that can be handled in our calculation.  Fig. \ref{theory}(a) and (b) show that both the condensate and thermal components oscillate during the free evolution period, and Fig. \ref{theory}(c) shows that the total number of atoms after the second dissipation period oscillates as a function of free evolution duration. Fig. \ref{theory}(d) plots the oscillation frequency and amplitudes as a function of trapping frequency, which shows the same trend as the experimental data shown in Fig. \ref{mechanism}(d).

\textit{Discussion and Outlook.} In summary, our work demonstrates that dissipation can be utilized as a tool to control quantum dynamics. We show that dissipation can alter the momentum distribution and effectively tune the system's temperature. Previously, to increase the temperature of ultracold atom systems, one usually applies a periodical moving local potential or periodically modulates optical lattices. These operations often introduce density or current excitations or excite high-energy modes, which prevents observing coherent exchange dynamics between the thermal and condensate components. Our method introduces neither density and current modes nor high-energy excitations, which is the major advantage responsible for our observations. The coherent dynamics between the thermal and condensate components are also related to the second sound of a superfluid \cite{stamper-kurn98,peshkov2013second,sidorenkov2013second}. Therefore, our method provides an alternative route to excite the second sound. With our method, it is also conceivable to observe dipole or quadrupole modes of thermal excitations, uncovering a novel and fundamental aspect of the superfluid.

\textit{Acknowledge.} This work is supported by the National Key R\&D Program of China 2023YFA1406702(WC and HZ), 
2021YFA1400904 (WC), 2021YFA0718303 (JH), and 2022YFA1405300 (YC), the Innovation Program for Quantum Science and Technology 2021ZD0302005 (HZ), the XPLORER Prize (HZ), NSFC Grant No. 92165203 (WC and JH), 12174358(YC), No. U23A6004 (HZ), and No. 12204352 (CW), Tsinghua University Initiative Scientific Research Program (JH,HZ,WC), and and NSAF (Grant No. U2330401)(YC).

\bibliography{references.bib}

\end{document}

\title{Supplementary Materials: \\Dissipation Driven Coherent Dynamics Observed in Bose-Einstein Condensates}
	
\author{Ye Tian}
\thanks{These authors contributed equally.}
\affiliation{Department of Physics and State Key Laboratory of Low Dimensional Quantum Physics, Tsinghua University, Beijing, 100084, China}
\author{Yajuan Zhao}
\thanks{These authors contributed equally.}
\affiliation{Department of Physics and State Key Laboratory of Low Dimensional Quantum Physics, Tsinghua University, Beijing, 100084, China}
\author{Yue Wu}
\thanks{These authors contributed equally.}
\affiliation{Institute for Advanced Study, Tsinghua University, Beijing, 100084, China}
\author{Jilai Ye}
\affiliation{Department of Physics and State Key Laboratory of Low Dimensional Quantum Physics, Tsinghua University, Beijing, 100084, China}
\author{Shuyao Mei}
\affiliation{Department of Physics and State Key Laboratory of Low Dimensional Quantum Physics, Tsinghua University, Beijing, 100084, China}
\author{Zhihao Chi}
\affiliation{Department of Physics and State Key Laboratory of Low Dimensional Quantum Physics, Tsinghua University, Beijing, 100084, China}
\author{Tian Tian}
\affiliation{Department of Physics and State Key Laboratory of Low Dimensional Quantum Physics, Tsinghua University, Beijing, 100084, China}
\author{Ce Wang}
\affiliation{School of Physics Science and Engineering, Tongji University, Shanghai, 200092, China}
\author{Zheyu Shi}
\affiliation{State Key Laboratory of Precision Spectroscopy, East China Normal University, Shanghai 200062, China}
\author{Yu Chen}
\email{ychen@gscaep.ac.cn}
\affiliation{Graduate School of China Academy of Engineering Physics, Beijing, 100193, China}
\author{Jiazhong Hu}
\email{hujiazhong01@ultracold.cn}
\affiliation{Department of Physics and State Key Laboratory of Low Dimensional Quantum Physics, Tsinghua University, Beijing, 100084, China}
\affiliation{Beijing Academy of Quantum Information Sciences, Beijing 100193, China}
\affiliation{Frontier Science Center for Quantum Information and Collaborative Innovation Center of Quantum Matter, Beijing, 100084, China}
\author{Hui Zhai}
\email{hzhai@tsinghua.edu.cn}
\affiliation{Institute for Advanced Study, Tsinghua University, Beijing, 100084, China}
\affiliation{Hefei National Laboratory, Hefei 230088, China}
\author{Wenlan Chen}
\email{cwlaser@ultracold.cn}
\affiliation{Department of Physics and State Key Laboratory of Low Dimensional Quantum Physics, Tsinghua University, Beijing, 100084, China}
\affiliation{Frontier Science Center for Quantum Information and Collaborative Innovation Center of Quantum Matter, Beijing, 100084, China}

\date{\today}

\maketitle
	
In this Supplementary Material, we first calculate different atom loss rates for the Bose-Einstein condensate and thermal gas. The results show consistency between the intensity of the dissipation light and atom loss rates and explain why the condensate loses faster than the thermal atoms.
After that, we build a two-fluid theory that includes the dynamics of Bose-Einstein condensates and thermal atoms in the presence of dissipation. We begin by reviewing the standard Bogoliubov theory, which helps us solve the initial equilibrium state at finite temperatures. Then, we introduce the normal and anomalous density matrices to describe the atom exchange between the condensates and thermal atoms. Finally, we construct a two-fluid dissipation theory based on the non-Hermitian linear response theory. This new dissipation theory allows us to theoretically predict the coherent oscillations between the condensates and thermal atoms, consistent with the experimental observations described in the main text.

\section{Calculation of different loss rates}
\begin{figure}[h]
    \centering
    \includegraphics[width=0.48\textwidth]{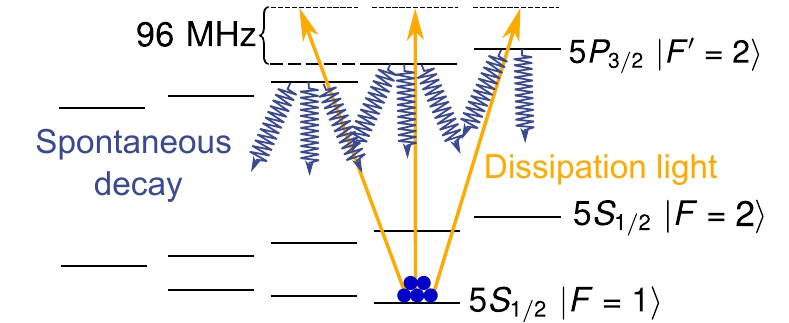}
    \caption{Illustration of the energy levels and dissipation light polarization. Atoms prepared in $5S_{1/2}\ket{F=1,m_F=1}$ are coupled to three different $5P_{3/2}$ levels, $\ket{F'=2,m_{F'}=0}$, $\ket{F'=2,m_{F'}=1}$, and $\ket{F'=2,m_{F'}=2}$ by $\sigma^-$, $\pi$, and $\sigma^+$ polarized light respectively. 
 Once atoms are excited into $5P_{3/2}$ levels,
they can decay to $5S_{1/2}\ket{F=2}$ and $5S_{1/2}\ket{F=1}$ through the spontaneous decay. 
The orange solid lines with arrows correspond to the coupling of dissipation light, and the blue wavy lines with arrows correspond to different spontaneous decay channels.
Each black solid line corresponds to a magnetic sublevel, and their positions illustrate the linear Zeeman shifts \cite{steck2001rubidium}.}
    \label{sm1}
\end{figure}

In our system, atoms are initialized in the $5S_{1/2}$ hyperfine ground state $\ket{F=1,m_F=1}$ of rubidium-87. These atoms are confined in a crossed dipole trap, and a magnetic gradient field $30\ \rm{G/cm}$ is applied to compensate the gravity so that only atoms in the states $\ket{F=1,m_F=1}$ and $\ket{F=2,m_F=-1}$ can be held in the trap. 
In order to introduce atom loss in the system, three pairs of counter-propagating dissipation light beams are incident onto the atoms. And this scheme helps us to avoid continuous momentum kicks along one direction. 
The frequency of dissipation light is chosen to be $96$ MHz~blue detuned to the transition of $5S_{1/2}\ket{F=1,m_F=1}$ $\rightarrow$ $5P_{3/2}\ket{F'=2,m_{F'}=1}$ in a $10\ \rm{Gauss}$ bias magnetic field along the $z$ direction. 
%Based on the main axis defined by this bias field, 
When we choose the $z$ axis as the quantization axis, 
we can decompose the dissipation light into three components: $\sigma^+$, $\sigma-$, and $\pi$. Details about the energy levels and dissipative light polarizations are shown in Fig.~\ref{sm1} for a better illustration.

Since the intensity of the dissipation light is small compared to the saturation intensity ($I/I_{sat}\sim0.01$), we may simplify the scattering process as follows. 
Atoms in $\ket{F=1,m_F=1}$ are coupled to $\ket{F'=2,m_{F'}=0}$, $\ket{F'=2,m_{F'}=1}$, or $\ket{F'=2,m_{F'}=2}$ by the different polarizations $\sigma^-$, $\pi$, or $\sigma^+$ respectively. 
The population of each excited state is $\Omega_s^2/4\Delta^2_s$ based on the perturbative theory, where $s$ denotes polarization of the light ($\sigma^-$, $\pi$, or $\sigma^+$), $\Omega_s$ is the Rabi frequency, and $\Delta_s$ is the detuning. 
Then the decay rate through each excited level is $\Gamma \Omega_s^2/4\Delta^2_s$ where $\Gamma$ is the spontaneous decay rate of excited states.
Because there are multiple states in $5S_{1/2}$, atoms can decay to different ground states with probabilities 
%with this population in excited states, spontaneous decay will cause atoms in excited states to decay to different ground sublevels with probabilities 
given by the Clebsch-Gordan coefficients \cite{steck2001rubidium}. 
In Table \ref{CG}, we list the probabilities in a final state $\ket{F,m_F}$ if one atom is initially in an excited state $\ket{F',m_{F'}}$ before the spontaneous decay. This can help us to calculate the scattering rates for each channel from the initial state $\ket{F=1,m_F=1}$ to a final state $\ket{F,m_F}$.
Here only atoms in $\ket{F=1,m_F=1}$ and $\ket{F=2,m_F=-1}$ can be held in the dipole trap, and atoms in the other states will quickly escape from the trap. And this is the dominant loss mechanism for thermal atoms.

\begin{table}
\centering
\begin{tabular}{|c|c|c|c|}

\hline
\diagbox{$5S_{1/2}$}{$5P_{3/2}$}&$\ket{F'=2,m_{F'}=0}$&$\ket{F'=2,m_{F'}=1}$&$\ket{F'=2,m_{F'}=2}$\\
\hline
$\ket{F=2,m_F=-2}$&0&0&0\\
\hline
$\ket{F=2,m_F=-1}$&$1/4$&0&0\\
\hline
$\ket{F=2,m_F=0}$&0&$1/4$&0\\
\hline
$\ket{F=2,m_F=1}$&${1}/{4}$&${1}/{12}$&$1/{6}$\\
\hline
$\ket{F=2,m_F=2}$&0&${1}/{6}$&${1}/{3}$\\
\hline
$\ket{F=1,m_F=-1}$&${1}/{12}$&0&0\\
\hline
$\ket{F=1,m_F=0}$&${1}/{3}$&${1}/{4}$&0\\
\hline
$\ket{F=1,m_F=1}$&${1}/{12}$&${1}/{4}$&${1}/{2}$\\
\hline

\end{tabular}
\caption{Probabilities of different final states $5S_{1/2}\ket{F,m_F}$ through an excited state  $5P_{3/2}\ket{F',m_{F'}}$. 
Here we assume one atom is initially in an excited state, either $\ket{F'=2,m_{F'}=0}$, $\ket{F'=2,m_{F'}=1}$, or $\ket{F'=2,m_{F'}=2}$. Once a spontaneous decay event happens, we consider how large 
 the probability is for this atom to decay into different ground states in the $5S_{1/2}$ manifold. 
These probabilities are calculated using the Clebsch-Gordan coefficients \cite{steck2001rubidium}.}
\label{CG}
\end{table}

Besides, we need to consider one more factor for the condensate. 
The recoil energy $E_k$ is in a regime $E_k\sim~0.1U_d$ and $E_k\sim40 T_c$, where $U_d$ is the depth of the dipole trap and $T_c$ denotes the temperature of condensates in Fig.~3a. 
Consequently, for a pure condensate, atoms that decay to $\ket{F=1,m_F=1}$ or $\ket{F=2,m_F=-1}$ after a spontaneous emission event will leave the condensate and become thermal atoms. 
This is an additional loss channel for the condensate.
Thus, the loss rates for a pure condensate and thermal gas are different. 

Utilizing the measured light intensities and frequencies for $\sigma^-$, $\pi$, and $\sigma^+$, we calculate the population $\Omega_s^2/4\Delta^2_s$ of each excited state, and the results are $P_0=6\times10^{-7}$, $P_1=2.9\times10^{-6}$, and $P_2=5.5\times10^{-6}$ for $\ket{F'=2,m_{F'}=0}$, $\ket{F'=2,m_{F'}=1}$, and $\ket{F'=2,m_{F'}=2}$ respectively. 
Based on these data, the rate at which atoms decay into different sublevels can be calculated by:
\begin{equation}
\Gamma_j=\sum_i \Gamma P_iD_{ji},
\end{equation}
where $\Gamma_j$ is the rate atoms decay to $j$th sublevel in $5S_{1/2}$, $P_i$ is the population of a excited state and $D_{ji}$ represents the matrix element in Table~\ref{CG}. So the rate for atoms decaying to
$\ket{F=1,m_F=1}$ and $\ket{F=2,m_F=-1}$ is $139\ \rm{s^{-1}}$, and the rate for atoms decaying to other sublevels is $201\ \rm{s^{-1}}$. 
Therefore, based on this qualitative model, the ratio of the loss rates of the thermal atoms and condensate is estimated as $0.59$, which is close to the measured data ($0.42$) shown in Fig. 3a. 
%This explains why the condensate and thermal atoms have different loss rates.

\section{Review of the Standard Bogoliubov theory}

First, we will have a brief review of the standard Bogoliubov theory \cite{dalfovo1999theory,pethick2008bose}. 
Here we choose a one-dimensional system for simplicity while it can still capture and describe all phenomena in the experimental setup.
%Here we take the one-dimensional system due to its simplicity and it is believed that the main mechanism can be captured in this way. 
We choose that the temperature of the system is very low but not zero. This means that both the condensates and the thermal atoms should be included in consideration. 

The Hamiltonian of interacting atoms at low temperatures is
\begin{equation}
    H = \int dx \hat{\Psi}^\dagger(x)\left(-\frac{1}{2m}\frac{\partial^2}{\partial x^2}+\frac{1}{2}m\omega^2x^2\right)\hat{\Psi}(x)+\frac{g}{2}\hat{\Psi}^\dagger(x)\hat{\Psi}^\dagger(x)\hat{\Psi}(x)\hat{\Psi}(x),
\end{equation}
where \(\hat{\Psi}(x)\) is a bosonic field operator of atoms, $m$ is the atomic mass, and $g$ is the s-wave contact interaction strength. Here we set the Planck constant $\hbar$ as unity. Then the field operator \(\hat{\Psi}(x,t)\) evolves under the Heisenberg equation with a form of
\begin{equation}
    i\frac{\partial}{\partial t}\hat{\Psi}(x,t)=\left(-\frac{1}{2m}\frac{\partial^2}{\partial x^2}+\frac{1}{2}m\omega^2x^2+g\hat{\Psi}^{\dagger}(x,t)\hat{\Psi}(x,t)\right)\hat{\Psi}(x,t).
    \label{eq2}
\end{equation}
To separate the condensate and the thermal atoms, we convert the bosonic field operator $\hat{\Psi}(x,t)$ into a form of
\begin{equation}
    \hat{\Psi}(x,t) = \Phi(x,t) + \hat{\psi}(x,t).
    \label{eq3}
\end{equation}
Here $\Phi(x,t)$ is a complex function defined as the expectation of the field operator 
$\Phi(x,t)=\langle\hat{\Psi}(x,t)\rangle$, where $\langle\cdot\rangle$ is the ensemble average. It is actually the condensate wave function and resembles the condensate part. $\hat{\psi}(x,t)$ is the quantum fluctuation away from the mean field description $\Phi(x,t)$. It can not be ignored at finite temperatures because it describes the thermal atoms. We need to notice that $\Phi$ is on the order of $\sqrt{N}$ and $\hat{\psi}$ is on the order of unity for such a partial-condensed system with a total atom number $N$. By counting the orders of $N$, the dynamics of $\Phi(x,t)$ and $\hat{\psi}(x,t)$ can be separated as followings

\begin{equation}
\begin{aligned}
    i\frac{\partial}{\partial t}\Phi(x,t)= &\left(-\frac{1}{2m}\frac{\partial^2}{\partial x^2}+\frac{1}{2}m\omega^2x^2\right)\Phi(x,t) 
    \\&+ g|\Phi(x,t)|^2\Phi(x,t) + 2g\langle\hat{\psi}(x,t)^{\dagger}\hat{\psi}(x,t)\rangle\Phi(x,t) + g\langle\hat{\psi}(x,t)\hat{\psi}(x,t)\rangle\Phi^{*}(x,t),
    \label{eq4}
    \end{aligned}
\end{equation}
\begin{equation}
\begin{aligned}
    i\frac{\partial}{\partial t}\hat{\psi}(x,t)=& \left(-\frac{1}{2m}\frac{\partial^2}{\partial x^2}+\frac{1}{2}m\omega^2x^2\right)\hat{\psi}
    (x,t) + 2g|\Phi(x,t)|^2\hat{\psi}(x,t) \\&+ g\Phi(x,t)^2\hat{\psi}^{\dagger}(x,t)+2g\langle\hat{\psi}^{\dagger}(x,t)\hat{\psi}(x,t)\rangle\hat{\psi}(x,t)+g\langle\hat{\psi}(x,t)\hat{\psi}(x,t)\rangle\hat{\psi}^{\dagger}(x,t).
    \label{eq5}
\end{aligned}
\end{equation}
Here the mean field approximations such as $\hat{\psi}^{\dagger} \hat{\psi}\hat{\psi}\approx 2\langle\hat{\psi}^{\dagger} \hat{\psi}\rangle\hat{\psi}+\langle\hat{\psi}\hat{\psi}\rangle\hat{\psi}^{\dagger}$ have been integrated to simplify the calculations.

\par Following the Bogoliubov approximation, we express $\hat{\Psi}(x,t)$ in terms of a superposition of quasi-particle operators {$\hat{a}_i$} and quasi-hole operators {$\hat{a}_i^{\dagger}$}
\begin{equation}
    \hat{\Psi}(x,t)=e^{-i\mu t}\left[\Phi(x) +\sum_i\left(u_i(x)e^{-i\epsilon_i t}\hat{a}_i-v_i^{*}(x)e^{i\epsilon_i t}\hat{a}_i^{\dagger}\right)\right],
    \label{eq6}
\end{equation}
where $u_i(x)$ and $v_i(x)$ are complex functions, $\mu$ is the chemical potential and \(\epsilon_i\) is the eigen-energy of the mode $i$. %Then putting everything together, it gives the spectrum of the system. 
By inserting Eq.~\ref{eq6} into Eq.~\ref{eq4} and Eq.~\ref{eq5}, we obtain
\begin{equation}
    \left(-\frac{1}{2m}\frac{\partial^2}{\partial x^2}+\frac{1}{2}m\omega^2x^2-\mu\right)\Phi(x) + g|\Phi(x)|^2\Phi(x) + 2g\langle\hat{\psi}^{\dagger}(x)\hat{\psi}(x)\rangle\Phi(x) + g\langle\hat{\psi}(x)\hat{\psi}(x)\rangle\Phi(x)=0,
\end{equation}
\begin{equation}
    \left(-\frac{1}{2m}\frac{\partial^2}{\partial x^2}+\frac{1}{2}m\omega^2x^2-\mu\right)u_i(x) + 2g|\Phi(x)|^2u _i(x)+2g\langle\hat{\psi}^{\dagger}(x)\hat{\psi}(x)\rangle u_i(x)-g\langle\hat{\psi}(x)\hat{\psi}(x)\rangle v_i(x)-g\Phi(x)^2v_i(x)=\epsilon_i u_i(x),
\end{equation}
\begin{equation}
    \left(-\frac{1}{2m}\frac{\partial^2}{\partial x^2}+\frac{1}{2}m\omega^2x^2-\mu\right)v_i(x) + 2g|\Phi(x)|^2v_i(x)+2g\langle\hat{\psi}^{\dagger}(x)\hat{\psi}(x)\rangle v_i(x)-g\langle\hat{\psi}(x)\hat{\psi}(x)\rangle^*u_i(x)-g\Phi^*(x)^{2}u_i(x)=-\epsilon_i v_i(x).
\end{equation}
Here the ensemble average $\langle\cdot\rangle$ is for a partial-condensed state, so there are equalities 
\begin{equation}
\langle\hat{\psi}^{\dagger}(x)\hat{\psi}(x)\rangle=\sum_iN_i\left(|u_i(x)|^2+|v_i(x)|^2\right)
\end{equation} 
and 
\begin{equation}
\langle\hat{\psi(x)}\hat{\psi(x)}\rangle=\sum_iN_i\left(-2u_i(x)v_i^*(x)\right), 
\end{equation}
where $N_i=\frac{1}{e^{\beta\epsilon_i}-1}$ is the occupation number of quasi-particles at temperature \(T=1/({\rm k_B}\beta)\) according to the Bose-Einstein distribution. From these self-consistent equations, we can solve out the stationary state solution $\Phi(x)$, $\epsilon_i$, $u_i(x)$, and $v_i(x)$. It is straightforward to see that it is not so convenient to calculate the non-equilibrium dynamics of both condensates and thermal atoms, therefore, we will try to develop another method to calculate the dynamics in the next section.

\section{A Two-Fluid Theory for condensates and thermal atoms}
In this section, we introduce a new approach to simulate the dynamical evolution of the mixture of condensates and thermal atoms.
This new approach can be easily generalized to open systems with dissipation, which we will show later. The key idea here is to establish a set of dynamic equations that can be presented by physical observables and their spatial derivatives up to the second order.
For convenience we define one-body density matrix $n(x',x,t)$ and anomalous density matrix $n_\text{s}(x',x,t)$ as
\begin{equation}
    n(x',x,t) = \langle \hat{\psi}^{\dagger}(x',t)\hat{\psi}(x,t)\rangle
\end{equation}
and
\begin{equation}
    n_\text{s}(x',x,t) = \langle \hat{\psi}(x',t)\hat{\psi}(x,t)\rangle.
\end{equation}
Here $\langle \cdot\rangle$ is the thermal ensemble average over the initial state. The diagonal elements of $n(x',x,t)$ are just the density of thermal atoms. The time-evolution equation of $n(x',x,t)$ can be derived via the dynamics of $\hat{\Psi}(x,t)$ with a form of
\begin{eqnarray}
i\partial_t n(x',x,t)&=&\left\langle \left(i\partial_t\hat{\psi}^\dag(x',t)\right)\hat{\psi}(x,t)+\hat{\psi}^\dag(x',t)\left(i\partial_t\hat{\psi}(x,t)\right)\right\rangle\nonumber\\
    &=&\left(-\frac{1}{2m}\frac{\partial^2}{\partial x^{2}}+\frac{1}{2}m\omega^2x^{2}+\frac{1}{2m}\frac{\partial^2}{\partial x'^{2}}-\frac{1}{2}m\omega^2x'^{2}+V_{\rm eff}^-[\Phi, n]\right)n(x',x,t)\nonumber\\
    &&+\left(g\Phi(x,t)^2+gn_\text{s}^{}(x,x,t)\right)n_\text{s}^*(x,x',t)-\left(g\Phi^{*}(x',t)^2+gn_\text{s}^*(x',x',t)\right)n_\text{s}(x',x,t),
\label{eq12}
\end{eqnarray}
where $V^-_{\rm eff}[\Phi,n]\equiv \left(2g|\Phi(x,t)|^2+2gn(x,x,t)-2g|\Phi(x',t)|^2-2gn(x',x',t)\right)$ is the effective potential generated by the condensates and thermal atoms. The same argument applies to $n_s(x',x,t)$ and we obtain
\begin{eqnarray}
i\partial_t n_\text{s}(x',x,t)&=&\left\langle \left(i\partial_t\hat{\psi}(x',t)\right)\hat{\psi}(x,t)+\hat{\psi}(x',t)\left(i\partial_t\hat{\psi}(x,t)\right)\right\rangle\nonumber\\
    &=&\left(-\frac{1}{2m}\frac{\partial^2}{\partial x^{2}}+\frac{1}{2}m\omega^2x^{2}-\frac{1}{2m}\frac{\partial^2}{\partial x'^{2}}+\frac{1}{2}m\omega^2x'^{2}+V^+_{\rm eff}[\Phi,n]\right)n_\text{s}(x',x,t)
\nonumber\\&&+\left(g\Phi(x,t)^2+gn_\text{s}(x,x,t)\right)n(x,x',t)+\left(g\Phi(x',t)^2+gn_\text{s}(x',x',t)\right)n(x',x,t),
\label{eq13}
\end{eqnarray}
where $V^+_{\rm eff}[\Phi,n]\equiv \left(2g|\Phi(x,t)|^2+2gn(x,x,t)+2g|\Phi(x',t)|^2+2gn(x',x',t)\right)$.
%, and we omit the $\delta(x-x')$ therm in the expression of $n(x,x',t)+\delta(x-x')$. Notice that $V^\pm_{\rm eff}[\Phi, n]$ are different in signs between the potential at position $x$ and $x'$.

Together with the equation of $\Phi(x,t)$,
\begin{eqnarray}
    i\frac{\partial}{\partial t}\Phi(x,t)&= &\left(-\frac{1}{2m}\frac{\partial^2}{\partial x^2}+\frac{1}{2}m\omega^2x^2\right)\Phi(x,t) + g|\Phi(x,t)|^2\Phi(x,t) + 2g n(x,x,t)\Phi(x,t) + g n_s(x,x,t)\Phi^{*}(x,t),
    \label{eq14}
\end{eqnarray}
Eq.~\ref{eq12}, Eq.~\ref{eq13}, and Eq.~\ref{eq14} form a set of self-consistent equations to solve the dynamics of condensate $\Phi(x,t)$, thermal density $n(x',x,t)$, anomalous density $n_s(x',x,t)$, and other associated parameters. 
By performing the Wigner transformation on $n(x',x,t)$ and $n_s(x',x,t)$, we can obtain hydrodynamical information $n(X,p,t)=\int n(x',x,t)e^{ip(x'-x)}d(x'-x)$ and $n_s(X,p,t)=\int n_s(x',x,t)e^{ip(x'-x)}d(x'-x)$ where $X=(x'+x)/2$. From these distributions $n(X,p,t)$ and $n_s(X,p,t)$ in the phase space, this method can resemble the Boltzmann equation. 

\section{A Two-fluid Theory with Dissipation}

\par In this section, we will extend the two-fluid theory to open quantum systems. When there is dissipation, a dissipative term \(\hat{H}_\text{diss}=\int dx\left(-i\gamma_\text{t}\hat{\Psi}^{\dagger}(x)\hat{\Psi}(x)+\hat{\Psi}^\dag(x)\xi(x)+\xi^\dag(x)\hat{\Psi}(x)\right)\) is introduced into the Hamiltonian of the system. Here we can set the Langevin noise operators satisfying $\langle \hat{\xi}(x,t)\hat{\xi}^\dag(x',t')\rangle=2\gamma_\text{t}\delta(t-t')\delta(x-x')$ for thermal atoms, and the change of \(n(x',x,t)\) due to the dissipation is
\begin{equation}
    \begin{aligned}
        &\delta\partial_tn(x',x,t)\\
        &= -\gamma_\text{t}\int dx''\langle\{\hat{\Psi}^\dag(x',t)\hat{\Psi}(x,t),\hat{\Psi}^\dag(x'',t)\hat{\Psi}(x'',t)\}-2\hat{\Psi}^\dag(x'',t)\hat{\Psi}^\dag(x',t)\hat{\Psi}(x,t)\hat{\Psi}(x'',t)\rangle\\
        &=-\gamma_\text{t} n(x',x,t),
    \end{aligned}
\end{equation}
according to the non-Hermitian linear response theory \cite{pan2020non}.
Similarly, the same argument can apply to $n_s(x',x,t)$ and we obtain the change of $n_s(x',x,t)$ in a form of
\begin{equation}
    \begin{aligned}
            &\delta\partial_t{n_\text{s}(x',x,t)}\\
        &= -\gamma_\text{t}\int dx''\langle\{\hat{\Psi}(x',t)\hat{\Psi}(x,t),\hat{\Psi}^\dag(x'',t)\hat{\Psi}(x'',t)\}-2\hat{\Psi}^\dag(x'',t)\hat{\Psi}(x',t)\hat{\Psi}(x,t)\hat{\Psi}(x'',t)\rangle\\
        &=-\gamma_\text{t} n_\text{s}(x',x,t).
    \end{aligned}
\end{equation}
Together with the evolution equations (Eq.~\ref{eq12} and Eq.~\ref{eq13}), we obtain two equations for $n(x',x,t)$ and $n_\text{s}(x',x,t)$ under dissipation,
\begin{eqnarray}
i\partial_t n(x',x,t)
    &=&\left(-\frac{1}{2m}\frac{\partial^2}{\partial x^{2}}+\frac{1}{2}m\omega^2x^{2}+\frac{1}{2m}\frac{\partial^2}{\partial x'^{2}}-\frac{1}{2}m\omega^2x'^{2}+V_{\rm eff}^-[\Phi, n]\right)n(x',x,t)\nonumber\\
    &&+\left(g\Phi(x,t)^2+gn_\text{s}(x,x,t)\right)n_\text{s}^*(x,x',t)-\left(g\Phi^{*}(x',t)^2+gn_\text{s}^*(x',x',t)\right)n_\text{s}(x',x,t) -i\gamma_\text{t} n(x',x,t)  
\label{eq16}
\end{eqnarray}
and
\begin{eqnarray}
i\partial_t n_\text{s}(x',x,t)
    &=&\left(-\frac{1}{2m}\frac{\partial^2}{\partial x^{2}}+\frac{1}{2}m\omega^2x^{2}-\frac{1}{2m}\frac{\partial^2}{\partial x'^{2}}+\frac{1}{2}m\omega^2x'^{2}+V_{\rm eff}^+[\Phi,n]\right)n_\text{s}(x',x,t)\nonumber\\
&&+\left(g\Phi(x,t)^2+gn_\text{s}(x,x,t)\right)n(x,x',t)+\left(g\Phi(x',t)^2+gn_\text{s}(x',x',t)\right)n(x',x,t) -i\gamma_\text{t} n_\text{s}(x',x,t).
\label{eq17}
\end{eqnarray}
However, the Bose-Einstein condensates do not satisfy the correlation assumption of the Langevin noise because the environment condenses at the same time.
% induces the condensing of environment and destroys our assumptions on the correlations of Langevin noise. Because of that, 
Then, we cannot apply the same argument to the dissipation of the condensate wave function $\Phi(x,t)$. So we introduce a different phenomenological dissipative parameter \(\gamma_c\) into the equation of $\Phi(x,t)$ and it leads to
\begin{eqnarray}
    i\frac{\partial}{\partial t}\Phi(x,t)&=& \left(-\frac{1}{2m}\frac{\partial^2}{\partial x^2}+\frac{1}{2}m\omega^2x^2\right)\Phi(x,t) + g|\Phi(x,t)|^2\Phi(x,t)  \nonumber\\
    &&+ 2g n(x,x,t)\Phi(x,t)+ gn_s(x,x,t)\Phi^{*}(x,t)-i\gamma_\text{c}\Phi(x,t).
\end{eqnarray}
Based on these equations above, we can solve how $\Phi(x,t)$ and $n(x,x,t)$ evolve with time, then predict the coherent oscillations between the condensates and the
thermal atoms which are consistent with the experimental observations.

\bibliography{SM_ref.bib}